# Broadcasting of Inseparability

Satyabrata Adhikari[a+], B.S.Choudhury[a], Indranil Chakrabarty[b]

[a]Department of Mathematics, Bengal Engineering College & Science University

Howrah-711103, West Bengal, India. [+]E-Mail: **satyyabrata@yahoo.com**

[b]Department of Mathematics, Heritage Institute of Technology,

Kolkata- 107, West Bengal, India.

**Abstract:** Suppose we are given an entangled pair and then one can ask how well we can produce two entangled pairs starting from a given entangled pair using only local operations. To give response of the above asked question, we study broadcasting of entanglement using state dependent quantum cloning machine as a local copier. We show that the length of the interval for probability-amplitude-squared $\alpha^2$ for broadcasting of entanglement using state dependent cloner can be made larger than the length of the interval for probability-amplitude-squared for broadcasting entanglement using state independent cloner. Further we show that there exists local state dependent cloner which gives better quality copy (in terms of average fidelity) of an entangled pair than the local universal cloner.

PACS Numbers: 03.67.-a, 03.65.Bz, 89.70.+c

Keywords: Broadcasting, Quantum cloning machine, Entanglement, Fidelity

**I. Introduction:**

Linearity of quantum theory prevents us from duplicating and deleting an unknown quantum state. Its consequence is the no-cloning theorem [1] and no-deletion theorem [15], which states that an ideal quantum copying machine and perfect quantum deleting machine does not exist. Although nature prevents us from amplifying an unknown quantum state but we can construct a quantum cloning machine that duplicates an unknown quantum state approximately [1,2,3,4,5]. Quantum copying machine can be divided into two classes: (a) Deterministic quantum copying machine and (b) Probabilistic quantum copying machine. The first type of quantum cloning machine can be further divided into two sub-classes: (i) State dependent quantum cloning machine, for example Wootters-Zurek (W-Z) quantum cloning machine [1], whose copying quality depends on the input state. (ii) Universal quantum copying machine, for example Buzek-Hillery (B-H) quantum cloning machine [2], whose copying quality remains same for all input state. In addition, the performance of universal B-H quantum cloning machine is, on average, better than that of the state dependent W-Z cloning machine. The fidelity of cloning of B-H universal quantum copying machine is $5/6$ which is better than any other existing universal quantum cloning machine. The latter type of quantum cloner i.e. Probabilistic quantum cloning machine clone an unknown quantum state, secretly chosen from a certain set of linearly independent states, accurately but with certain probabilities less than unity [16].

Entanglement [14] is a quantum mechanical feature that can be employed for computational and communicational purposes. Therefore, as a valuable resource in quantum information processing, quantum entanglement has been widely used in quantum cryptography [10,13], quantum superdense coding [11] and quantum teleportation [12]. Consequently, it remains the subject of interest at present after years of investigations. Among all the problems regarding entanglement, broadcasting of entanglement is an important issue to consider. Broadcasting is nothing but a local copying of non-local quantum correlations. In this process, the entanglement originally shared by two observers is broadcast into two identical less entangled states by using a local $1 \rightarrow 2$ optimal universal symmetric cloning machine.

**Definition:**

Suppose two distant parties A and B share two qubit-entangled state

$|s\rangle_{AB} = \alpha|00\rangle_{AB} + \beta|11\rangle_{AB}$ with $\alpha^2 + |\beta|^2 = 1$.

The first qubit belongs to A and the second belongs to B. Each of the two parties now perform local copier on their own qubit and then the input entangled state $|\psi\rangle$ has been broadcast if for some values of the probability $\alpha^2$

(1) non-local output states are inseparable, and

(2) local output states are separable.

The above-described process was used by Buzek et.al. [6] for broadcasting entanglement using Universal quantum cloning machine as a local copier. Broadcasting (cloning) of non-local correlations of quantum states also studied by S.Bandyopadhyay et.al. [9] and showed that broadcasting of more than two pairs from a single pair is not possible using local copier. In the process of broadcasting of entanglement, we generally use Peres-Horodecki theorem for showing the inseparability of non-local outputs and separability of local outputs.

**Peres-Horodecki Theorem [7,8]:** The necessary and sufficient condition for the state $\hat{\rho}$ of two spins ½ to be inseparable is that at least one of the eigen values of the partially transposed operator defined as $\rho^{T_2}_{m\mu,n\nu} = \rho_{m\nu,n\mu}$ is negative. This is equivalent to the condition that at least one of the two determinants

$$W_3 = \begin{vmatrix} \rho_{00,00} & \rho_{01,00} & \rho_{00,10} \\ \rho_{00,01} & \rho_{01,01} & \rho_{00,11} \\ \rho_{10,00} & \rho_{11,00} & \rho_{10,10} \end{vmatrix} \text{ and } W_4 = \begin{vmatrix} \rho_{00,00} & \rho_{01,00} & \rho_{00,10} & \rho_{01,10} \\ \rho_{00,01} & \rho_{01,01} & \rho_{00,11} & \rho_{01,11} \\ \rho_{10,00} & \rho_{11,00} & \rho_{10,10} & \rho_{11,10} \\ \rho_{10,01} & \rho_{11,01} & \rho_{10,11} & \rho_{11,11} \end{vmatrix}$$

is negative and

$$W_2 = \begin{vmatrix} \rho_{00,00} & \rho_{01,00} \\ \rho_{00,01} & \rho_{01,01} \end{vmatrix} \text{ is non-negative.}$$

Now, we distribute our work in the remaining three sections. In section II, we introduce a state dependent quantum-cloning machine, which we will use in the broadcasting process later. In section III, we revisit the broadcasting of entanglement procedure proposed by Buzek et.al. In section IV, we discuss the broadcasting of entanglement via state dependent cloning machine and show that broadcasting is possible in a wider range of the probability $\alpha^2$ compared with the range of the probability for broadcasting of entanglement via universal cloning machine as a local copier.

**II. State dependent B-H quantum cloning machine:**

In the literature, many state dependent quantum cloners were known. In this section, we also introduce another state dependent cloner. The introduced state dependent cloner is interesting in the sense that it can be constructed from B-H quantum cloning transformation by relaxing one condition of universality viz. $\partial D_{ab}/\partial \alpha^2 = 0$ where $D_{ab} = Tr[\rho^{(out)}_{ab} - \rho^{(id)}_a \otimes \rho^{(id)}_b]^2$. $\rho^{(out)}_{ab}$ describes the entangled output states of the cloner and $\rho^{(id)}_a$, $\rho^{(id)}_b$ describes the input state in mode 'a' and 'b' respectively.

The B-H cloning transformation is given by

$$|0\rangle|\Sigma\rangle|Q\rangle \rightarrow |0\rangle|0\rangle|Q_0\rangle + (|0\rangle|1\rangle + |1\rangle|0\rangle)|Y_0\rangle \qquad (2.1)$$

$$|1\rangle|\Sigma\rangle|Q\rangle \to |1\rangle|1\rangle|Q_1\rangle + (|0\rangle|1\rangle + |1\rangle|0\rangle)|Y_1\rangle \qquad (2.2)$$

The unitarity of the transformation gives

$$\langle Q_i|Q_i\rangle + 2\langle Y_i|Y_i\rangle = 1, \quad i = 0,1 \qquad (2.3)$$

$$\langle Y_0|Y_1\rangle = \langle Y_1|Y_0\rangle = 0 \qquad (2.4)$$

We assume

$$\langle Q_0|Y_0\rangle = \langle Q_1|Y_1\rangle = \langle Q_1|Q_0\rangle = 0 \qquad (2.5)$$

Let $|\psi\rangle = \alpha|0\rangle + \beta|1\rangle$ \qquad (2.6)

with $\alpha^2 + |\beta|^2 = 1$, be the input state.

We assume $\alpha$ is real and $\beta$ is complex.

The cloning transformation (2.1-2.2) copy the information of the input state (2.6) approximately into two identical states described by the density operators $\rho_a^{(out)}$ and $\rho_b^{(out)}$ respectively.

The output state described by the density operator $\rho_b^{(out)}$ looks the same as $\rho_a^{(out)}$.

The reduced density operator $\rho_a^{(out)}$ is given by

$$\rho_a^{(out)} = |0\rangle\langle 0| \left[\alpha^2 + \left(|\beta|^2 \langle Y_1|Y_1\rangle - \alpha^2 \langle Y_0|Y_0\rangle\right)\right] + |0\rangle\langle 1|\alpha\beta^* \left[\langle Q_1|Y_0\rangle + \langle Y_1|Q_0\rangle\right]$$

$$+ |1\rangle\langle 0|\alpha\beta \left[\langle Q_1|Y_0\rangle + \langle Y_1|Q_0\rangle\right] + |1\rangle\langle 1|\left[|\beta|^2 - \left(|\beta|^2 \langle Y_1|Y_1\rangle - \alpha^2 \langle Y_0|Y_0\rangle\right)\right]$$

$$= |0\rangle\langle 0|\left[\alpha^2 + \lambda\left(|\beta|^2 - \alpha^2\right)\right] + |0\rangle\langle 1|\alpha\beta^* \mu + |1\rangle\langle 0|\alpha\beta \mu + |1\rangle\langle 1|\left[|\beta|^2 - \lambda\left(|\beta|^2 - \alpha^2\right)\right] \qquad (2.7)$$

where $\langle Y_0|Y_0\rangle = \langle Y_1|Y_1\rangle = \lambda$ \qquad (2.8)

$$\langle Q_0|Y_1\rangle = \langle Q_1|Y_0\rangle = \langle Y_1|Q_0\rangle = \langle Y_0|Q_1\rangle = \mu/2 \qquad (2.9)$$

The distortion of the qubit in mode 'a' is

$$D_a = 2\lambda^2 \left(4\alpha^4 - 4\alpha^2 + 1\right) + 2\alpha^2 \left(1 - \alpha^2\right)(\mu - 1)^2 \qquad (2.10)$$

The distortion $D_{ab}$ is defined by

$$D_{ab} = Tr[\rho_{ab}^{(out)} - \rho_a^{(id)} \otimes \rho_b^{(id)}]^2$$

$$= Tr\begin{bmatrix} U_{11} & U_{12} & U_{13} \\ U_{21} & U_{22} & U_{23} \\ U_{31} & U_{32} & U_{33} \end{bmatrix}^2$$

$$= U_{11}^2 + 2|U_{12}|^2 + 2|U_{13}|^2 + U_{22}^2 + 2|U_{23}|^2 + U_{33}^2 \qquad (2.11)$$

where $U_{11} = \alpha^4 - \alpha^2(1-2\lambda)$ \qquad (2.12a)

$$U_{12} = \sqrt{2}\alpha^3\beta^* - \sqrt{2}\alpha\beta^*\mu/2 \,, \, U_{21} = (U_{12})^* \qquad (2.12b)$$

$$U_{13} = \alpha^2(\beta^*)^2 \,, \, U_{31} = (U_{13})^* \qquad (2.12c)$$

$$U_{22} = 2\alpha^2|\beta|^2 - 2\lambda \qquad (2.12d)$$

$$U_{23} = \sqrt{2}\alpha\beta^*|\beta|^2 - \sqrt{2}\alpha\beta^*\mu/2 \,, \, U_{32} = (U_{23})^* \qquad (2.12e)$$

$$U_{33} = |\beta|^4 - |\beta|^2(1-2\lambda) \qquad (2.12f)$$

The cloning transformation (2.1-2.2) is input state independent if $D_a$ and $D_{ab}$ are input state independent. In this work, we are interested in input state dependent cloning machine. To make the cloning transformation (2.1-2.2) input state dependent, we assume $D_{ab}$ is input state dependent i.e. $\partial D_{ab}/\partial\alpha^2 \neq 0$. \qquad (2.13)

The relation between the machine parameters $\lambda$ and $\mu$ is established by solving the equation $\partial D_a/\partial\alpha^2 = 0$. Therefore, $\partial D_a/\partial\alpha^2 = 0 \Rightarrow \mu = 1 - 2\lambda$. \qquad (2.14)

The value of the machine parameter $\lambda$ is restricted from the condition $\partial D_{ab}/\partial\alpha^2 \neq 0$. The above condition (2.13) implies that $\lambda$ can take any value between 0 and $1/2$ except $1/6$. However, if $\lambda = 1/6$, then $\partial D_a/\partial\alpha^2 = 0$ and $\partial D_{ab}/\partial\alpha^2 = 0$, therefore the machine becomes universal in the sense that it does not depend on the input state.

Putting $\mu = 1 - 2\lambda$ in (2.11) & (2.12a-2.12f), we get

$$D_{ab} = [\alpha^4 - \alpha^2(1-2\lambda)]^2 + 4\alpha^2(1-\alpha^2)(\alpha^2 - (1-2\lambda)/2)^2 + 2\alpha^4(1-\alpha^2)^2 + +(2\alpha^2(1-\alpha^2) - 2\lambda)^2$$
$$+ 4\alpha^2(1-\alpha^2)(1-\alpha^2 - (1-2\lambda)/2)^2 + (1-\alpha^2)^2(2\lambda - \alpha^2)^2 \qquad (2.15)$$

For maximum or minimum value of $D_{ab}$, we have

$$\partial D_{ab}/\partial\lambda = 0 \Rightarrow \lambda = 3\alpha^2(1-\alpha^2)/4 \qquad (2.16)$$

Again, $\partial^2 D_{ab}/\partial\lambda^2 = 16 > 0$ \qquad (2.17)

Equation (2.17) implies that $D_{ab}$ has minimum value when the machine parameter $\lambda$ takes the form given in equation (2.16).

Thus we are able to construct a quantum-cloning machine where machine state vectors depends on input state and therefore the quality of the copy depends on the input state i.e. for different input states, machine state vectors take different values and hence the quality of the copy changes.

Putting $\mu = 1 - 2\lambda$ in (2.10), we get

$D_a(\alpha^2) = 2\lambda^2$, Since $\lambda$ depends on $\alpha^2$. (2.18)

TABLE-1

| Probability ($\alpha^2$) | For State Dependent cloner | | For B-H State Independent Cloner | |
|---|---|---|---|---|
| | Machine Parameter $\lambda = 3\alpha^2(1-\alpha^2)/4$ | Distance Between Input and Output State, $D_a = 2\lambda^2$ | Machine Parameter $\lambda = 1/6$ | Distance Between Input and Output State, $D_a$ |
| 0.1 | 0.007 | 0.000098 | 0.167 | 0.055556 |
| 0.2 | 0.029 | 0.001682 | 0.167 | 0.055556 |
| 0.3 | 0.061 | 0.007442 | 0.167 | 0.055556 |
| 0.4 | 0.101 | 0.020402 | 0.167 | 0.055556 |
| 0.5 | 0.141 | 0.039762 | 0.167 | 0.055556 |
| 0.6 | 0.173 | 0.059858 | 0.167 | 0.055556 |
| 0.7 | 0.187 | 0.069938 | 0.167 | 0.055556 |
| 0.8 | 0.173 | 0.059858 | 0.167 | 0.055556 |
| 0.9 | 0.115 | 0.026450 | 0.167 | 0.055556 |

The above table shows that the quality of the copy depends on the input state if we consider the B-H state dependent cloner while we can observe that the quality of the copy of B-H state independent cloner remains same for all input states

Finally we have constructed a state dependent quantum-cloning machine that we use for broadcasting of entanglement in section IV.

**III. Revisit the Broadcasting of entanglement**

In this section, we revisit the broadcasting of entanglement procedure by Buzek et.al.
Let the input entangled state be given by

$$|\phi\rangle = \alpha_1|00\rangle_{AB} + \beta_1|11\rangle_{AB} \qquad (3.1)$$

with real $\alpha_1$ and $\beta_1$ and $\alpha_1^2 + \beta_1^2 = 1$.

The state (3.1) is inseparable for all values of $\alpha_1^2$ such that $0 < \alpha_1^2 < 1$ because one of the two determinants $W_3$ and $W_4$ is negative and $W_2$ is non-negative.

Using the state independent universal B-H cloning machine as a local copier, the local output described by the density operator

$$\rho_{AA'} = \rho_{BB'} = 2\alpha_1^2/3\,|00\rangle\langle 00| + 1/3|+\rangle\langle +| + 2\beta_1^2/3|11\rangle\langle 11| \qquad (3.2)$$

where $|+\rangle = (1/\sqrt{2})(|01\rangle + |10\rangle)$

while the non-local output described by the density operator

$$\rho_{AB'} = \rho_{A'B} = (24\alpha_1^2 + 1)/36\,|00\rangle\langle 00| + (24\beta_1^2 + 1)/36|11\rangle\langle 11| + 5/36(|01\rangle\langle 01| + |10\rangle\langle 10|)$$
$$+ 4\alpha_1\beta_1/9\,(|00\rangle\langle 11| + |11\rangle\langle 00|) \qquad (3.3)$$

From Peres-Horodecki criteria for separability, it follows that $\rho_{AA'}\,(\rho_{BB'})$ is separable if

$$1/2 - \sqrt{48}/16 \le \alpha_1^2 \le 1/2 + \sqrt{48}/16 \qquad (3.4)$$

and $\rho_{AB'}\,(\rho_{A'B})$ is inseparable if

$$1/2 - \sqrt{39}/16 \le \alpha_1^2 \le 1/2 + \sqrt{39}/16 \qquad (3.5)$$

Therefore, the entanglement is broadcasted via local state independent quantum cloner if the probability- amplitude-squared $\alpha_1^2$ is given by the range

$$1/2 - \sqrt{39}/16 \le \alpha_1^2 \le 1/2 + \sqrt{39}/16. \qquad (3.6)$$

The fidelity of broadcasting is given by

$$F_1(\alpha_1^2) = \langle\phi|\rho_{AB'}|\phi\rangle = 25/36 - 4\alpha_1^2(1-\alpha_1^2)/9 \qquad (3.7)$$

From equation (3.7), we note that although the state independent cloner is used as a local cloner for broadcasting entanglement but we find that the fidelity of copying an entanglement depends on the input state. Thus, the actions of state independent cloner on the respective particles hold by two distant parties locally does not clone the entanglement equally for all values of the probability $\alpha_1^2$.

Hence, the average fidelity is given by

$$\overline{F}_1 = \int_0^1 F_1(\alpha_1^2) d\alpha_1^2 = 67/108 = 0.62 \tag{3.8}$$

## IV. Broadcasting of entanglement using state dependent B-H quantum cloning machine:

In this section, our aim is to show that the broadcasting of inseparability using state dependent quantum cloning machine locally is more effective than using state independent B-H quantum cloning machine.

Let us consider a general pure entangled state

$$|\chi\rangle_{AB} = \alpha_1|00\rangle + \beta_1|11\rangle + \gamma_1|10\rangle + \delta_1|01\rangle \tag{4.1}$$

where $\alpha_1, \beta_1, \gamma_1, \delta_1$ is real and $\alpha_1^2 + \beta_1^2 + \gamma_1^2 + \delta_1^2 = 1$.

The first qubit (A) belongs to Alice and the second qubit (B) belongs to Bob. Then the two distant partners Alice and Bob apply their respective state dependent quantum cloner on their qubits to produce two output systems $A'$ and $B'$ respectively. Now our task is to see whether local cloning procedure generates two pair of entanglement from a given entangled pair. Therefore, to investigate the existence of non-local correlations in two systems described by the non-local density operators $\{(\rho_{AB'}, \rho_{A'B})$ or $(\rho_{AB}, \rho_{A'B'})\}$, we use Peres-Horodecki criteria. Also, to test the separability of the local outputs described by the density operators $(\rho_{AA'}, \rho_{BB'})$, we use the same criteria as before.

The two non-local output states of a copier are described by the density operator $\rho_{AB'}$ & $\rho_{A'B}$,

$\rho_{AB'} = \rho_{A'B} = C_{11}|00\rangle\langle00| + C_{44}|11\rangle\langle11| + C_{22}|01\rangle\langle01| + C_{33}|10\rangle\langle10| +$
$C_{23}|00\rangle\langle11| + C_{23}|11\rangle\langle00| + C_{12}|01\rangle\langle00| + C_{12}|00\rangle\langle01| + C_{13}|00\rangle\langle10| +$

$C_{13}|10\rangle\langle00| + C_{14}|01\rangle\langle10| + C_{14}|10\rangle\langle01| + C_{24}|01\rangle\langle11| + C_{24}|11\rangle\langle01| + C_{34}|11\rangle\langle10| +$

$C_{34}|10\rangle\langle11|$ (4.2)

Where

$C_{11} = \alpha_1^2(1-\lambda)^2 + \beta_1^2\lambda^2 + \lambda(1-\lambda)(\delta_1^2 + \gamma_1^2)$ (4.3a)

$C_{12} = \beta_1\gamma_1\lambda\mu + \delta_1\alpha_1\mu(1-\lambda),\ C_{13} = \beta_1\delta_1\lambda\mu + \alpha_1\gamma_1\mu(1-\lambda),\ C_{14} = \mu^2\delta_1\gamma_1$ (4.3b)

$C_{22} = \delta_1^2(1-\lambda)^2 + \gamma_1^2\lambda^2 + \lambda(1-\lambda)(\alpha_1^2 + \beta_1^2)$ (4.3c)

$C_{23} = C_{32} = \mu^2\alpha_1\beta_1,\ C_{24} = \alpha_1\gamma_1\lambda\mu + \beta_1\delta_1\mu(1-\lambda)$ (4.3d)

$C_{33} = \gamma_1^2(1-\lambda)^2 + \delta_1^2\lambda^2 + \lambda(1-\lambda)(\alpha_1^2 + \beta_1^2),\ C_{34} = \delta_1\alpha_1\mu\lambda + \beta_1\gamma_1\mu(1-\lambda)$ (4.3e)

$C_{44} = \alpha_1^2\lambda^2 + \beta_1^2(1-\lambda)^2 + \lambda(1-\lambda)(\delta_1^2 + \gamma_1^2)$ (4.3f)

The two local output states of a copier are described by the density operators $\rho_{AA'}$ & $\rho_{BB'}$,

$\rho_{AA'} = K_{11}|00\rangle\langle00| + K_{44}|11\rangle\langle11| + K_{22}|01\rangle\langle01| + K_{33}|10\rangle\langle10| + K_{14}|01\rangle\langle10| + K_{41}|10\rangle\langle01|$

$+ K_{12}|01\rangle\langle00| + K_{12}|00\rangle\langle01| + K_{13}|00\rangle\langle10| + K_{13}|10\rangle\langle00| + K_{24}|01\rangle\langle11| + K_{24}|11\rangle\langle01| +$

$K_{34}|11\rangle\langle10| + K_{34}|10\rangle\langle11|$ (4.4)

Where $K_{11} = (1-2\lambda)(\alpha_1 + \delta_1)^2$ (4.5a)

$K_{12} = K_{13} = K_{24} = K_{34} = \left(\dfrac{\mu}{2}\right)(\alpha_1 + \delta_1)(\beta_1 + \gamma_1)$ (4.5b)

$K_{14} = K_{41} = K_{22} = K_{33} = \lambda + 2\lambda(\beta_1\gamma_1 + \delta_1\alpha_1)$ (4.5c)

$K_{23} = K_{32} = 0,\ K_{44} = (1-2\lambda)(\beta_1 + \gamma_1)^2$ (4.5d)

$\rho_{BB'} = K'_{11}|00\rangle\langle00| + K'_{44}|11\rangle\langle11| + K'_{22}|01\rangle\langle01| + K'_{33}|10\rangle\langle10| + K'_{14}|01\rangle\langle10|$

$+ K'_{41}|10\rangle\langle01| + K'_{12}|01\rangle\langle00| + K'_{12}|00\rangle\langle01| + K'_{31}|00\rangle\langle10| + K'_{31}|10\rangle\langle00| + K'_{42}|01\rangle\langle11| +$

$K'_{42}|11\rangle\langle01| + K'_{34}|11\rangle\langle10| + K'_{34}|10\rangle\langle11|$ (4.6)

Where $K'_{11} = (1-2\lambda)(\alpha_1 + \gamma_1)^2$ (4.6a)

$K'_{12} = K'_{31} = K'_{42} = K'_{34} = \left(\dfrac{\mu}{2}\right)(\alpha_1 + \gamma_1)(\beta_1 + \delta_1)$ (4.6b)

$K'_{14} = K'_{41} = K'_{22} = K'_{33} = \lambda + 2\lambda(\alpha_1\gamma_1 + \delta_1\beta_1)$ (4.6c)

$K'_{23} = K'_{32} = 0,\ K'_{44} = (1-2\lambda)(\beta_1 + \delta_1)^2$ (4.6d)

The composite systems described by the density operator $\rho_{AB'}$ and $\rho_{A'B}$ is inseparable if at least one of the determinants $W_3$ and $W_4$ is negative and $W_2$ is non-negative, where

$$W_3 = \begin{vmatrix} C_{11} & C_{12} & C_{13} \\ C_{12} & C_{22} & C_{23} \\ C_{13} & C_{23} & C_{33} \end{vmatrix},\ W_4 = \begin{vmatrix} C_{11} & C_{12} & C_{13} & C_{14} \\ C_{12} & C_{22} & C_{23} & C_{24} \\ C_{13} & C_{23} & C_{33} & C_{34} \\ C_{14} & C_{24} & C_{34} & C_{44} \end{vmatrix},\ W_2 = \begin{vmatrix} C_{11} & C_{12} \\ C_{12} & C_{22} \end{vmatrix} \quad (4.7)$$

The entries in the determinants are given by the equations (4.3a-4.3f).

The local output state in Alice's Hilbert space described by the density operator $\rho_{AA'}$ is separable if

$$W_3 = \begin{vmatrix} K_{11} & K_{12} & K_{13} \\ K_{12} & K_{22} & K_{23} \\ K_{13} & K_{32} & K_{33} \end{vmatrix} \geq 0,\ W_4 = \begin{vmatrix} K_{11} & K_{12} & K_{13} & K_{14} \\ K_{12} & K_{22} & K_{23} & K_{24} \\ K_{13} & K_{32} & K_{33} & K_{34} \\ K_{41} & K_{24} & K_{34} & K_{44} \end{vmatrix} \geq 0,\ W_2 = \begin{vmatrix} K_{11} & K_{12} \\ K_{12} & K_{22} \end{vmatrix} \geq 0 \quad (4.8)$$

The entries in the determinants are given by the equations (4.5a-4.5d).

The local output state in Bob's Hilbert space described by the density operator $\rho_{BB'}$ is separable if

$$W_3 = \begin{vmatrix} K'_{11} & K'_{12} & K'_{31} \\ K'_{12} & K'_{22} & K'_{23} \\ K'_{31} & K'_{32} & K'_{33} \end{vmatrix} \geq 0,\ W_4 = \begin{vmatrix} K'_{11} & K'_{12} & K'_{31} & K'_{14} \\ K'_{12} & K'_{22} & K'_{23} & K'_{42} \\ K'_{31} & K'_{32} & K'_{33} & K'_{34} \\ K'_{41} & K'_{42} & K'_{34} & K'_{44} \end{vmatrix} \geq 0,\ W_2 = \begin{vmatrix} K'_{11} & K'_{12} \\ K'_{12} & K'_{22} \end{vmatrix} \geq 0 \quad (4.9)$$

The entries in the determinants are given by the equations (4.6a-4.6d).

Now we say that the broadcasting is possible for general pure entangled state (4.1) if the equations (4.7-4.9) are satisfied.

For simplicity and without any loss of generality, we assume that the two distant parties Alice and Bob share a pair of particles prepared in the pure entangled state

$$|\chi\rangle = \alpha_1 |00\rangle_{AB} + \beta_1 |11\rangle_{AB} \quad (4.10)$$

where $\alpha_1$ is real and $\beta_1$ is a complex number such that $\alpha_1^2 + |\beta_1|^2 = 1$.

Alice and Bob then apply the state dependent quantum cloner as a local copier on their qubits. As a result, the two non-local output states of a copier are described by the density operators $\rho_{AB'}$ & $\rho_{A'B}$ and two local output states are described by the density operators $\rho_{AA'}$ and $\rho_{BB'}$.

The non-local density operators $\rho_{AB'}$ & $\rho_{A'B}$ are given by

$$\rho_{AB'} = \rho_{A'B} = |00\rangle\langle00|[\alpha_1^2(1-2\lambda)+\lambda^2] + \lambda(1-\lambda)(|01\rangle\langle01|+|10\rangle\langle10|) + |11\rangle\langle11|[|\beta_1|^2(1-2\lambda)+\lambda^2]$$
$$+\alpha_1\beta_1^*\mu^2|00\rangle\langle11|+\alpha_1^*\beta_1\mu^2|11\rangle\langle00| \qquad (4.11)$$

It follows from the Peres-Horodecki theorem that $\rho_{AB'}$ & $\rho_{A'B}$ are inseparable if

$$W_4 = \begin{vmatrix} (1-2\lambda)\alpha_1^2 & 0 & 0 & 0 \\ 0 & \lambda(1-\lambda) & \alpha_1\beta_1^*\mu^2 & 0 \\ 0 & \alpha_1^*\beta_1\mu^2 & \lambda(1-\lambda) & 0 \\ 0 & 0 & 0 & |\beta_1|^2(1-2\lambda)+\lambda^2 \end{vmatrix} < 0$$

$$\Rightarrow \alpha_1^4\mu^4 - \alpha_1^2\mu^4 + \lambda^2(1-\lambda)^2 < 0$$

$$\Rightarrow 1/2 - \left(\sqrt{\mu^4 - 4\lambda^2(1-\lambda)^2}/2\mu^2\right) < \alpha_1^2 < 1/2 + \left(\sqrt{\mu^4 - 4\lambda^2(1-\lambda)^2}/2\mu^2\right)$$

$$\Rightarrow 1/2 - \left(\sqrt{(1-2\lambda)^4 - 4\lambda^2(1-\lambda)^2}/2(1-2\lambda)^2\right) < \alpha_1^2 < 1/2 + \left(\sqrt{(1-2\lambda)^4 - 4\lambda^2(1-\lambda)^2}/2(1-2\lambda)^2\right)$$

Also we note that $W_3 < 0$ and $W_2 \geq 0$.

The local density operators $\rho_{AA'}$ & $\rho_{BB'}$ are given by

$$\rho_{AA'} = \rho_{BB'} = |00\rangle\langle00|\alpha_1^2(1-2\lambda) + \lambda(|01\rangle\langle01|+|10\rangle\langle10|+|01\rangle\langle10|+|10\rangle\langle01|)$$
$$+|11\rangle\langle11||\beta_1|^2(1-2\lambda) \qquad (4.12)$$

Now $\rho_{AA'}$ & $\rho_{BB'}$ are separable if $W_2 \geq 0, W_3 \geq 0$ and $W_4 \geq 0$.

$$W_4 = \begin{vmatrix} (1-2\lambda)\alpha_1^2 & 0 & 0 & \lambda \\ 0 & \lambda & 0 & 0 \\ 0 & 0 & \lambda & 0 \\ \lambda & 0 & 0 & (1-2\lambda)|\beta_1|^2 \end{vmatrix} \geq 0$$

$$\Rightarrow \alpha_1^4(1-2\lambda)^2 - \alpha_1^2(1-2\lambda)^2 + \lambda^2 \leq 0$$

$$\Rightarrow 1/2 - \sqrt{1-4\lambda}/2(1-2\lambda) \leq \alpha_1^2 \leq 1/2 + \sqrt{1-4\lambda}/2(1-2\lambda) \qquad (4.13)$$

TABLE-2

| Machine parameter, $\lambda$ | Interval ($I_1$) for Inseparability between systems (A-B$'$) or (A$'$-B) | Interval ($I_2$) for separability between Systems (A-A$'$) or (B-B$'$) | Common Interval between ($I_1$) and ($I_2$) |
|:---:|:---:|:---:|:---:|
| 0.007 | $(0.00005, 0.99994)$ | $(0.00005, 0.99994)$ | $(0.00005, 0.99994)$ |
| 0.029 | $(0.00101, 0.99899)$ | $(0.00094, 0.99905)$ | $(0.00101, 0.99899)$ |
| 0.061 | $(0.00555, 0.99444)$ | $(0.00485, 0.99514)$ | $(0.00555, 0.99444)$ |
| 0.101 | $(0.02076, 0.97923)$ | $(0.01628, 0.98371)$ | $(0.02076, 0.97923)$ |
| 0.115 | $(0.03038, 0.96961)$ | $(0.02282, 0.97717)$ | $(0.03038, 0.96961)$ |
| 0.141 | $(0.05863, 0.94136)$ | $(0.04017, 0.95982)$ | $(0.05863, 0.94136)$ |
| 0.159 | $(0.09091, 0.90908)$ | $(0.05768, 0.94231)$ | $(0.09091, 0.90908)$ |
| 0.173 | $(0.12836, 0.87163)$ | $(0.07570, 0.92429)$ | $(0.12836, 0.87163)$ |
| 0.187 | $(0.18458, 0.81541)$ | $(0.09904, 0.90095)$ | $(0.18458, 0.81541)$ |

Table-2 shows the interval for probability $\alpha_1^2$ for broadcasting of entanglement using state dependent quantum-cloning machine. Also we note from the above table that for the last two cases, the length of the intervals for broadcasting via state dependent cloner are smaller than the length of the interval for broadcasting discussed by Buzek et.al. while the situation is opposite in the remaining cases.

Now to see how well the local state dependent quantum cloners produce two entangled pairs from a single pair, we have to calculate the amount of overlapping between the input entangled state and the output entangled state described by the density operator $\rho_{AB'}$ ($\rho_{A'B}$).

Thus, The fidelity of broadcasting of inseparability is given by

$$F(\alpha_1^2) = \langle \chi | \rho_{AB'} | \chi \rangle = (1-\lambda)^2 - 4\alpha_1^2(1-\alpha_1^2)\lambda(1-2\lambda) \qquad (4.14)$$

The average fidelity is

$$\overline{F} = \int_0^1 F(\alpha_1^2) \, d\alpha_1^2 = (7\lambda^2 - 8\lambda + 3)/3 \qquad (4.15)$$

Now we are in a position to compare the techniques for broadcasting of entanglement using state dependent and state independent cloner.

(i) In the first technique, Buzek et.al used state independent cloner as a local copier but in the present technique, we use state dependent cloner as a local copying machine for broadcasting entanglement.

(ii) In the first technique where state independent quantum cloner was used, the broadcasting is possible in the interval $(0.10968, 0.89031)$ for $\alpha_1^2$ while in the second technique where state dependent quantum cloning machine is used as a local copier for broadcasting, the interval for the probability $\alpha_1^2$ depends on the machine parameter $\lambda$. Furthermore, we find that when the machine parameter takes the value lying in the interval $(0, 0.159]$, the interval for $\alpha_1^2$ in the second technique is much wider than the interval for $\alpha_1^2$ in the first technique. The situation becomes opposite when the value of the machine parameter $\lambda$ lying in the interval $[0.166, 0.5)$ i.e. in this case, the first technique dominates over the second. Table-2 supports the validity of the above statement. Therefore, we can observe that there exists some state dependent cloner with which the entanglement can be broadcasted with a wider range than the broadcasting entanglement using state independent cloner.

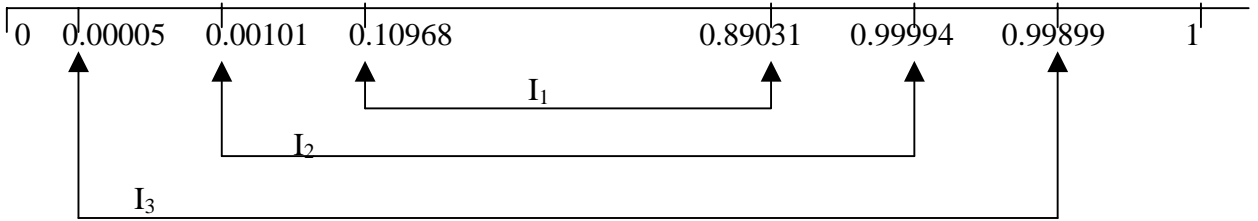

Figure: In the above figure, we show some intervals for inseparability between systems (A-B′) or (A′-B) to compare our technique with Buzek et.al. techniques for broadcasting entanglement. 'I$_1$' represent the interval for inseparability between systems when optimal universal quantum cloning machine is used as a local copier while 'I$_2$' and 'I$_3$' represents the interval for inseparability between systems when state dependent quantum cloning machine is used as a local copier.

(iii) The average fidelity of copying an entanglement is greater when using state dependent quantum cloner than the average fidelity of copying entanglement using state independent cloner when

$$\frac{(7\lambda^2 - 8\lambda + 3)}{3} > \frac{67}{108} \Rightarrow 0 < \lambda < 0.167 \text{ Or } 0.976 < \lambda < 1 \qquad (4.16)$$

We reject $0.976 < \lambda < 1$, since the machine parameter $\lambda$ lying between 0 and $\frac{1}{2}$.

Therefore, from table-2 we observe that there exist values of $\lambda$ for which broadcasting of entanglement and average fidelity of copying both can be performed better.

In summary, we have constructed a state dependent quantum cloning machine using B-H cloning transformation by relaxing one of the universality conditions then we studied state dependent quantum-cloning machine with different machine parameter $\lambda$ and the results are given in the table-1. Also we generalize the qubit states in the form $\alpha_1|00\rangle + \beta_1|11\rangle + \gamma_1|10\rangle + \delta_1|01\rangle$ and then studied the broadcasting of entanglement of the generalized pure state using the newly constructed local state-dependent quantum cloning machine. We also give the interval of the broadcasting of entanglement for different machine parameter $\lambda$ and these results are shown in table-2. Next, we consider a specific pure state of the form $\alpha_1|00\rangle + \beta_1|11\rangle$, which is also taken by Buzek et.al. and then comparing the two methods (using state dependent and state independent quantum cloning machine) for broadcasting of entanglement of the given pure entangled state. Next we have shown that the broadcasting of entanglement using state dependent quantum cloning machine has many advantages over the partial cloning of entanglement using state independent B-H quantum cloning machine.

The advantages are: (i) the length of the interval for $\alpha^2$ is broader in the case of broadcasting of inseparability using state dependent quantum cloning machine.

(ii) We get the better quality of copy on average of an entangled pair when we use state dependent quantum cloner than using state independent cloner locally.

**Acknowledgement:**

S.A gratefully acknowledge the partial support by CSIR under the project F.No.8/3(38)/2003-EMR-1, New Delhi. Authors also gratefully acknowledge the referees for their helpful comments.